\documentclass[12pt,twoside]{article}
\usepackage{fleqn,espcrc1}

\usepackage{graphicx}
\usepackage{epsfig}
\usepackage[figuresright]{rotating}

\newcommand{\AmS}{{\protect\the\textfont2
  A\kern-.1667em\lower.5ex\hbox{M}\kern-.125emS}}

\title{Recent N$^*$ Results From $J/\psi$ Decays}
\author{Haibo LI
        \thanks{e-mail address: lihb@hpws6.ihep.ac.cn}\\
        BES Collaboration\\
        H.C.Chiang, G.X.Peng and B.S.Zou\\
        Institute of High Energy Physics, P.O.Box 918(1), Beijing 100039\\
        P.R.China}
\begin{document}
\maketitle
\begin{abstract}
Based on 7.8 million $J/\psi$ events collected at BEPC,
the events for $J/\psi \to p\bar{p}\pi^0$ and $p\bar{p}\eta$ have
been selected and reconstructed. Clear peaks are observed around 1480 MeV 
in $p\pi^0 (\bar{p}\pi^0)$ invariant mass spectrum and near the threshold for
$\eta$ production in $p\eta(\bar{p}\eta)$ invariant mass spectrum. 
A partial wave analysis of $J/\psi \to p\bar{p}\eta$ data 
has been performed. Two $J^P=\frac{1}{2}^-$ resonances are observed
with mass and width (M, $\Gamma$) at ($1540^{+15}_{-17}$, $178^{+20}_{-22}$)
MeV and ($1648^{+18}_{-16}$, $150$)MeV, and are considered to be the nucleon
resonances $S_{11}(1535)$ and $S_{11}(1650)$ respectively.
\end{abstract}
\section{Introduction}

Nucleons are the most common form of hadronic matter on the earth
and probably in the whole universe. understanding their internal
structure will give us insight into how the real world works.
An important source of information about the nucleon internal structure
is the nucleon excitation spectrum. Our present knowledge on this aspect
came almost entirely from partial-wave analyses of $\pi N$ total, elastic,
and charge-exchange scattering data of more than twenty years
ago\cite{pdg}.
Since the late 1970's, very little has happened in experimental $N^*$
baryon spectroscopy. Considering its importance for the understanding
of the baryon structure and for distinguishing various pictures
\cite{isgur} of the nonperturbative regime of QCD, a new generation
of experiments on N* physics with electromagnetic probes have recently
been started at new facilities, such as CEBAF at JLAB, ELSA at Bonn,
GRAAL at Grenoble and so on.

One of us, Zou suggested \cite{zou} that we can also study $N^*$ baryon in 
J$/\psi$ decay to baryon-antibaryon final states, which provide a new
laboratory for study of $N^*$ baryon, especially in the mass range of 
1-2 GeV. For example, the $J/\psi \to p \bar{p} \eta$ is an excellent
channel to study the $N^*(1535)$ state which has a very large decay
branching ratio to the $N \eta $
\cite{kru,ben}, while other baryon resonances below 2.0 GeV do not have a 
large branching ratio to decay in this channel\cite{bar}, a fact noted 
very early in the development of the quark shell model \cite{mit}. 

In this paper, based on 7.8 million $J/\psi$ events collected at BEPC,
the events for $J/\psi \to p\bar{p}\pi^0$ and $p\bar{p}\eta$ have
been selected and reconstructed. We perform a partial wave analysis(PWA) 
on $J/\psi \rightarrow p \bar{p} \eta$ data in the full mass region of
$p\eta$($\bar{p} \eta$). This is the first PWA study of $N^*$ baryon
in $J/\psi$ hadronic decay in the world. Two S-wave $N^*$ baryon, namely
$N^*(1535)$ and $N^*(1650)$, are found in their $p \eta$ decay mode. 
The new information on $J/\psi N N^*$ couplings provides a new source for
studying baryon structure.  
\section{Event Selection}

The $\eta$ and $\pi^0$ are detected in their $\gamma\gamma$ decay modes.
Each candidate event is required to have two oppositely signed charged
tracks with a good helix fit in the polar
angle range $-0.8 < \cos\theta < 0.8$ in MDC and at least 2 reconstructed
$\gamma$'s in BSC. A vertex is required within an interaction region
$\pm 15$ cm longitudinally and 2 cm radially.   
A minimum energy cut of 60~MeV is imposed on the photons. 
Showers associated with charged tracks are also removed.

\begin{figure}[htb]
\begin{minipage}[t]{55mm}
\centerline{\epsfig{file=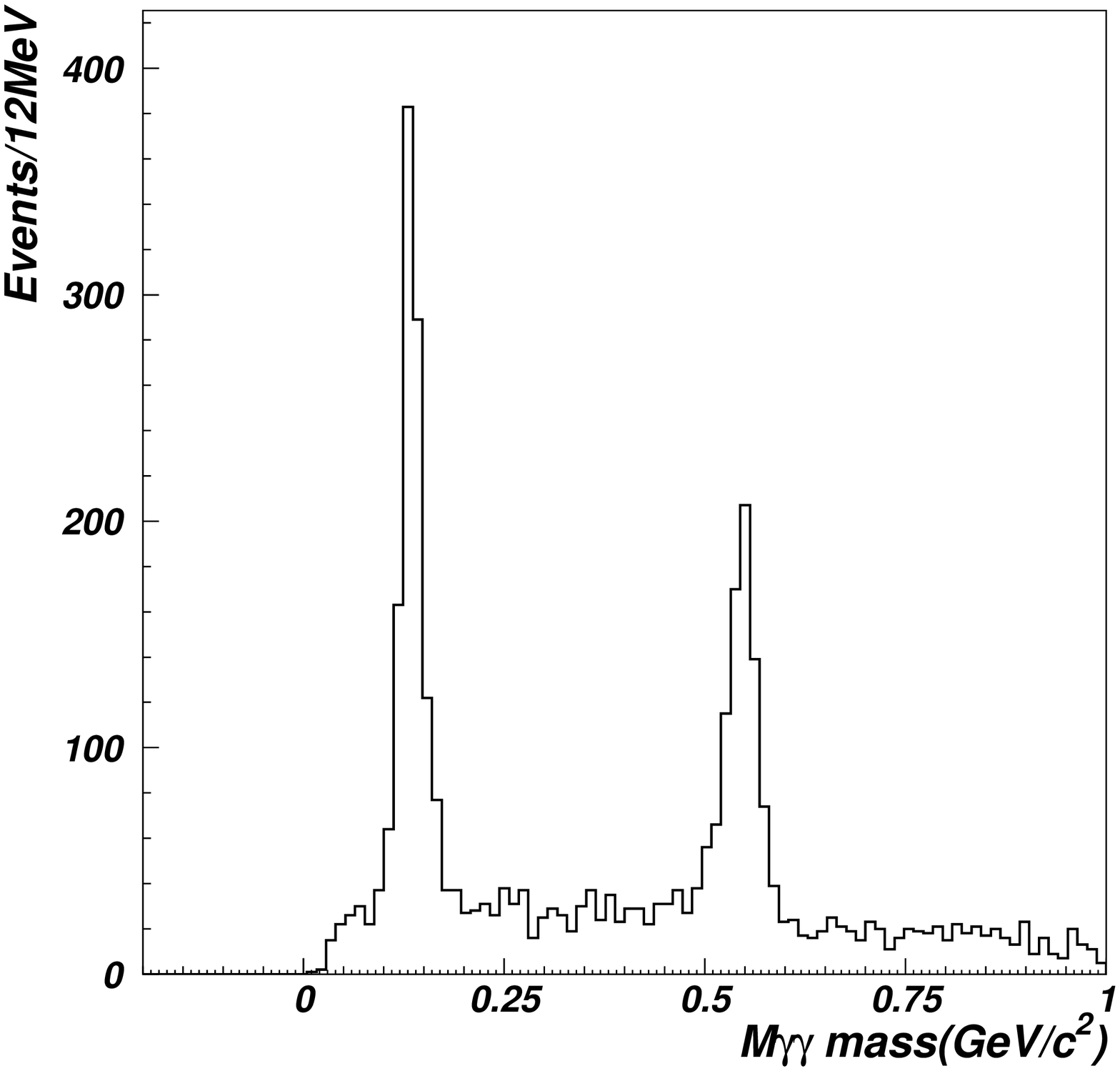,height=1.8in,width=2.0in}}
\caption{The Inv. mass of $\gamma \gamma$ after 4C fit}
\label{fig:gg}
\end{minipage}
\hspace{\fill}
\begin{minipage}[t]{50mm}
\centerline{\epsfig{file=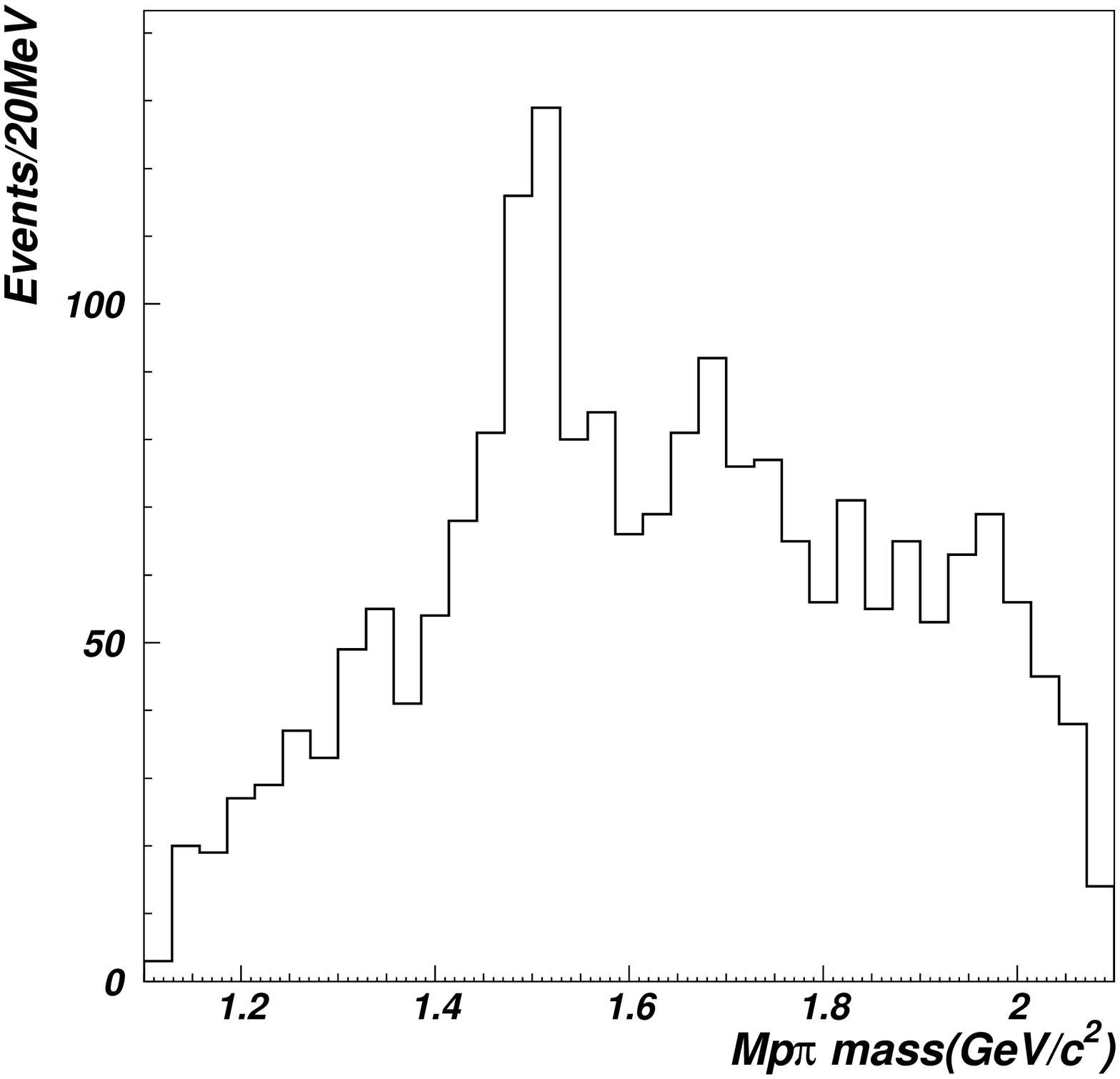,height=1.8in,width=2.0in}}
\caption{Mass spectrum for $p\pi^0$}
\label{fig:ppi0}
\end{minipage}
\hspace{\fill}
\begin{minipage}[t]{45mm}
\centerline{\epsfig{file=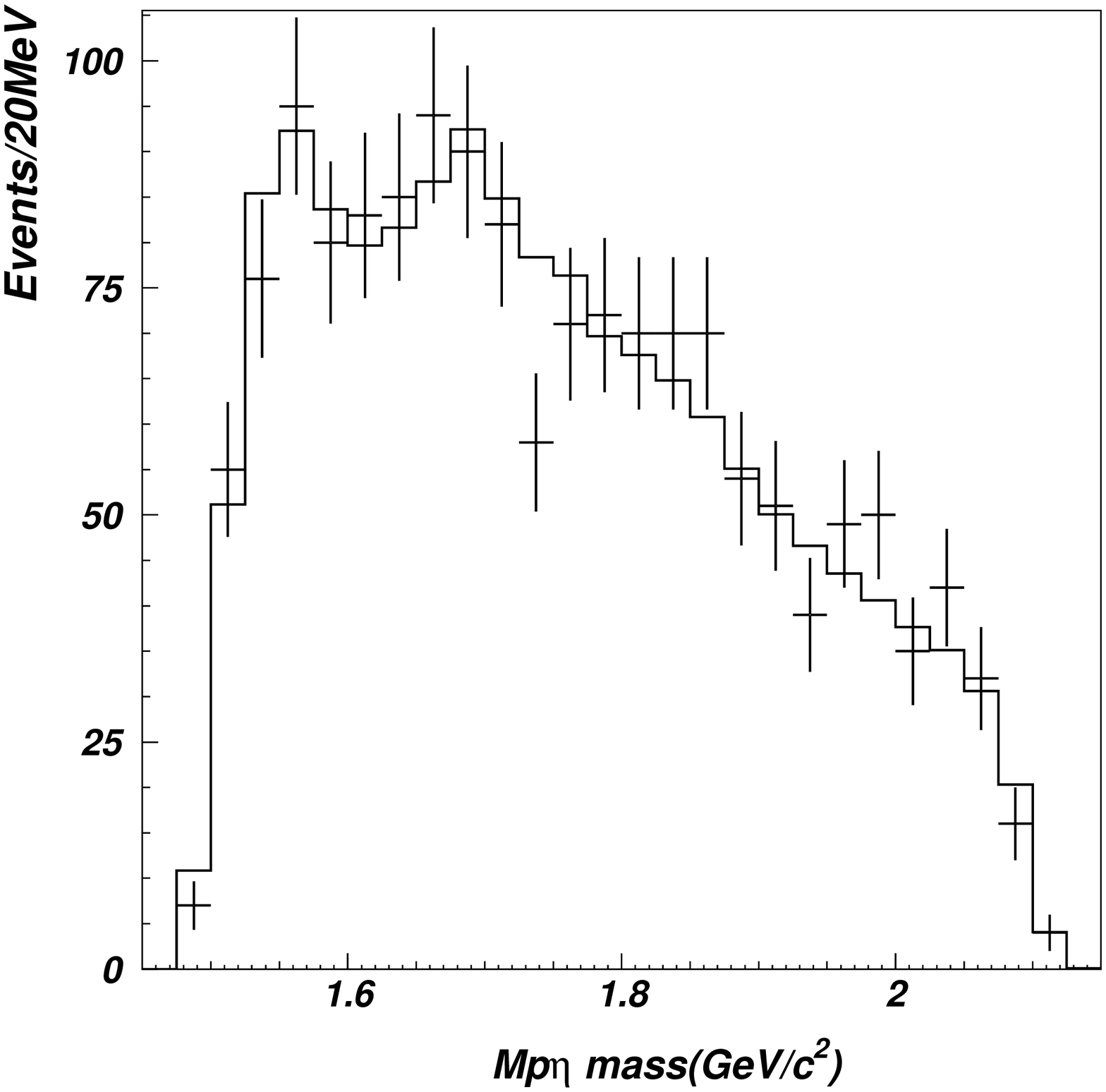,height=1.8in,width=2.0in}}
\caption{Mass spectrum for $p\eta$, Crosses are data and histograms the fit}
\label{fig:peta}
\end{minipage}
\end{figure}

After previous selection, we use TOF information to identify the $p\bar{p}$
pairs, and at least one track with unambiguous TOF information is required.
The open angle of two charged tracks smaller than $175^o$ is required in order
to remove back to back events; to remove radiative Bhabha events, we require
$(E_+/P_+ -1)^2+(E_-/P_- -1)^2 >0.4$, where $E_+$, $P_+$ ($E_-$, $P_-$)
are the energy deposited in BSC and momentum of positron (electron)
respectively. Events are fitted kinematically to the 4C hypotheses
$J/\psi \to 2\gamma p\bar{p}$. Figure ~\ref{fig:gg} shows the 
invariance mass spactrum of the $2\gamma$, we can see the clear $\pi^0$
and $\eta$ signals. Meanwhile, the events are also fitted to
$J/\psi \to \gamma p\bar{p}$ and $4\gamma p \bar{p}$. We require 
$$ Prob(\chi^2_{(2\gamma  p\bar{p})}, 4C) > Prob(\chi^2_{(\gamma  p\bar{p})}
, 4C),
~~~ Prob(\chi^2_{(2\gamma  p\bar{p})}, 4C) > Prob(\chi^2_{(4\gamma
p\bar{p})},4C)$$ 
to reject the $\gamma  p\bar{p}$ and $p\bar{p}\pi^0\pi^0$ backgrounds.
In order to improve the mass resolution, 5C fits are
performed on the selected events, the extra constraints are those of the 
$\eta$ and $\pi^0$ masses for $J/\psi \to p\bar{p}\eta$ and $p\bar{p}\pi^0$
decays, respectively. $Prob(\chi^2_{p\bar{p}\eta},5C) > 1\%$
($Prob(\chi^2_{p\bar{p}\pi^0},5C) > 1\%$) is required.

Figure ~\ref{fig:ppi0} and Figure ~\ref{fig:peta} show the $p\pi^0$ and 
$p\eta$ mass distributions from the decays $J/\psi \to p\bar{p}\pi^0$, 
$p\bar{p}\eta$ respectively. Clear peaks are observed around 1480 MeV 
in $p\pi^0$ invariant mass spectrum. The $p\eta$ events peak
strongly in the neighborhood of $\eta$-production threshold,
and we shall show that the data require a strong $\frac{1}{2}^-$ peak
near the threshold. There is an additional obvious bump
around 1600-1700MeV, it favors $J^P=\frac{1}{2}^-$ in our $S_{11}$
($\eta p$) partial wave analysis. 
\section{Amplitude Analysis of $J/\psi \to p\bar{p}\eta$}

A PWA analysis is performed for the $J/\psi \to p\bar{p}\eta$ channel\cite{li}
with the amplitudes constructed from Lorentz-invariant 
combinations of the momenta and the photon polarization 4-vectors for
$J/\psi$ initial states with helicity $\pm 1$. 
The relative magnitudes and phases of the amplitudes are determined by a
maximum likelihood fit to the data. Based on the study of $p\bar{p}$
and $p\eta$ invariant mass distributions in our data, the decay
chain $J/\psi \to p\bar{p}\eta$ is analyzed taking into account
two $p\eta$($\bar{p}\eta$) S-waves($S_{11}$) and one $p\eta$($\bar{p}\eta$) 
P-wave($P_{11}$) intermediate processes.  
The two S-waves amplitudes are only used to fit the data in the
low mass region near the $\eta$ production threshold. While the P-wave
is used to fit the data in the high mass region.
The background from multi-$\pi^0$ is $\sim ~8\%$ in the 5C fit, we 
have included a phase space background in the PWA fit to allow for this; 
The $p\eta$ mass projection fitted to the real data is shown
in Figure ~\ref{fig:peta}. 
We now discuss the features of the data and the outcome of fits.
\begin{figure}[htb]
\begin{minipage}[t]{55mm}
\centerline{\epsfig{file=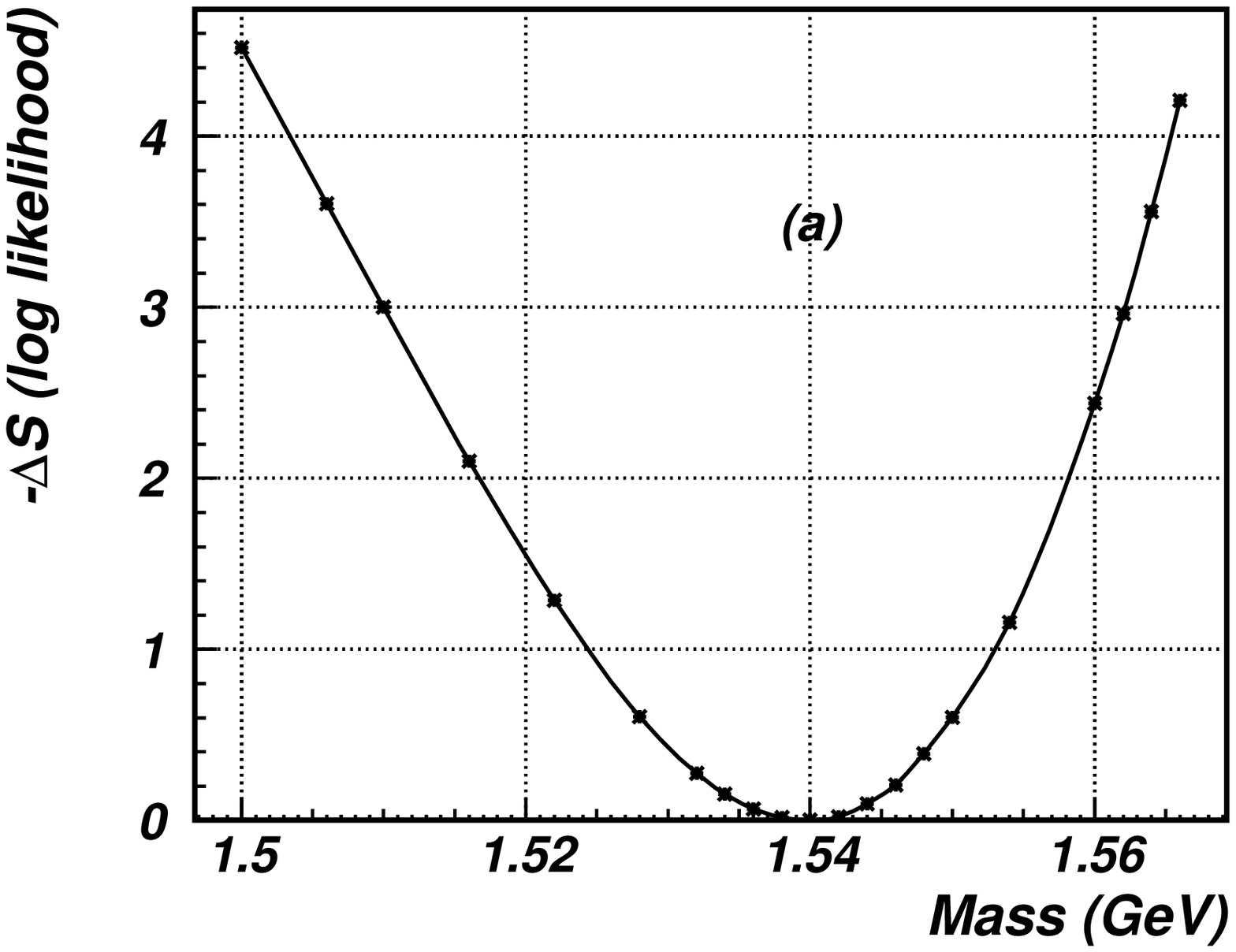,height=1.6in,width=2.0in}}
\end{minipage}
\hspace{\fill}
\begin{minipage}[t]{50mm}
\centerline{\epsfig{file=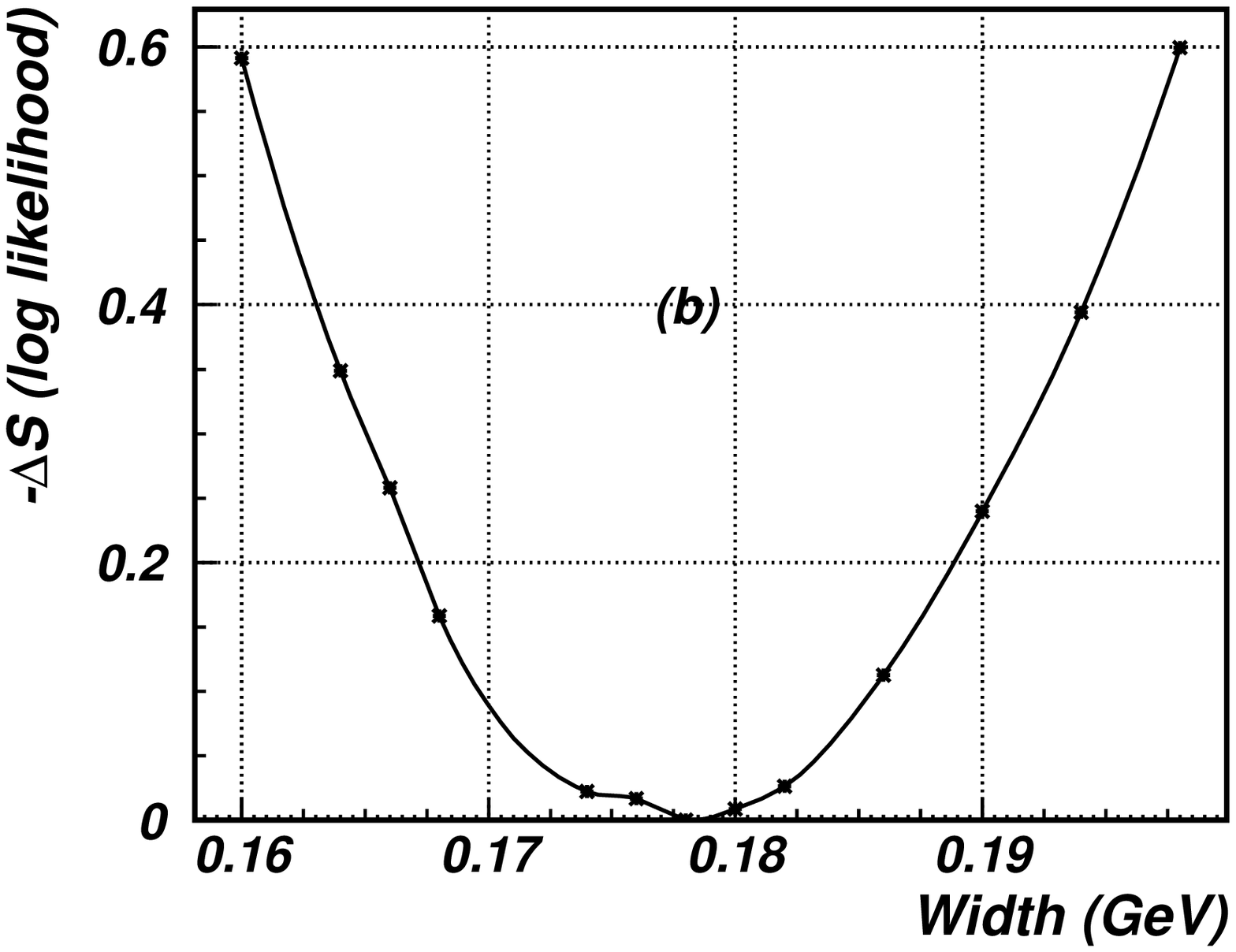,height=1.6in,width=2.0in}}
\end{minipage}
\hspace{\fill}
\begin{minipage}[t]{45mm}
\centerline{\epsfig{file=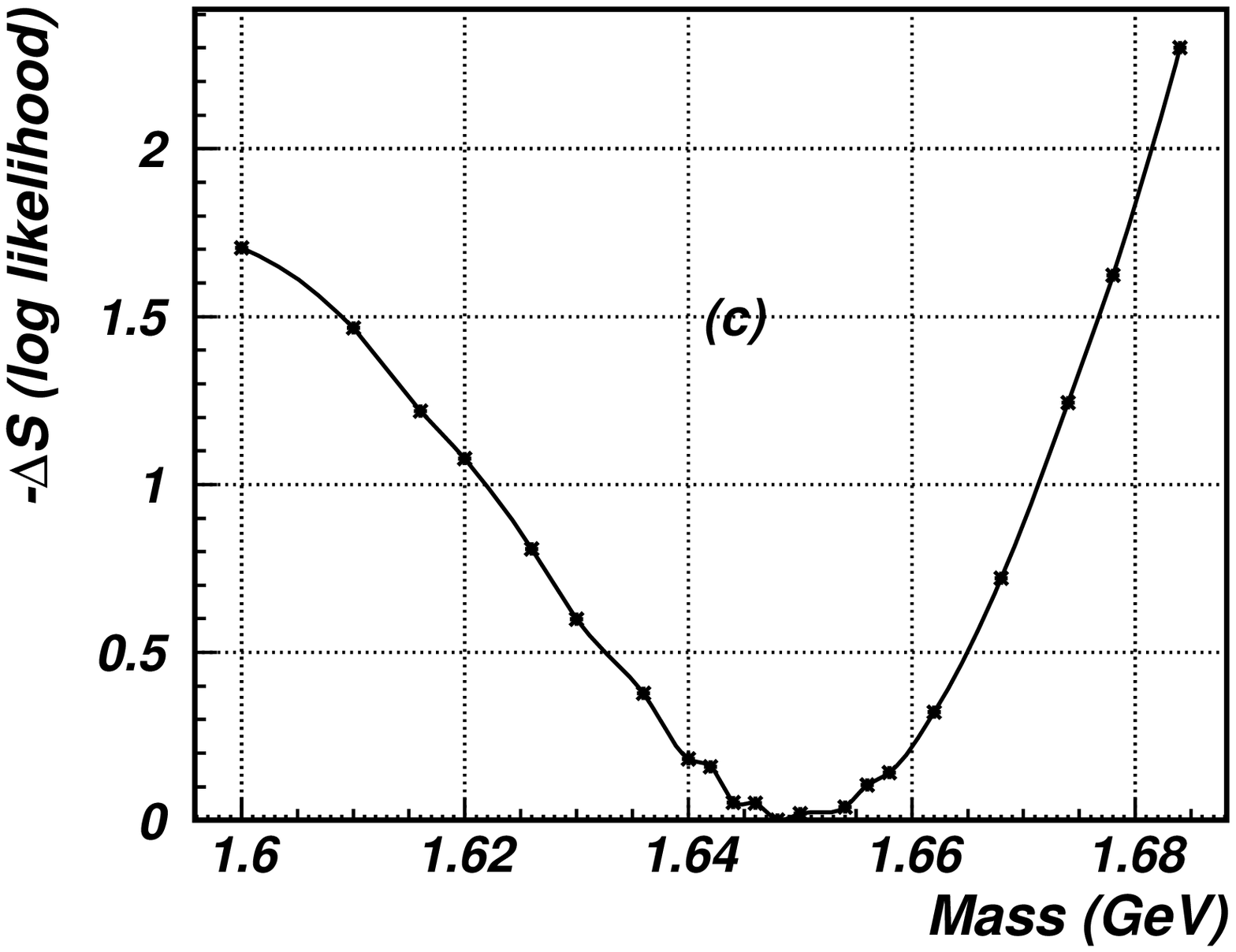,height=1.6in,width=2.0in}}
\end{minipage}
\caption[]{Scans for (a)the mass of $N^*(1535)$, (b)the width of
$N^*(1535)$ and (c)the mass of $N^*(1650)$} 
\label{fig:fit}
\end{figure}
\subsection{$S_{11}(1535)$}

A peak at $\sim$ 1535 MeV near the $\eta$ threshold optimises $M=1540
^{+15}_{-17}$ MeV as shown in Figure ~\ref{fig:fit}(a), The data favour 
$J^{P} =\frac{1}{2}^{-}$ over $ \frac{1}{2}^{+}$. A fit with $J^{P} =  
\frac{1}{2}^{+}$ instead gives $\ln L$ worse by 16.0 
than for $\frac{1}{2}^{-}$ assignment(Our definition of $log ~L$ is such
that it increases by 0.5 for a one standard deviation change in one 
parameter). With our 4 fitted parameters, the statistical significance
of the peak is above 6.0$\sigma$.  For the width scan as shown in Figure
~\ref{fig:fit}(b), our data require a width, $\Gamma = 178^{+20}_{-22}$ MeV.
Our results for $N^*(1535)$ are consistent with the resonance parameters
measured by Krusche et al.\cite{kru} at the MAMI acceletator in Mainz on
the $\eta$ photoproduction.

\subsection{$S_{11}(1650)$}

At $\sim $ 1650 MeV, there is a further peak. We fit it with a
$J^{P}=\frac{1}{2}^{-}$ resonance. Figure ~\ref{fig:fit}(c) are the scan of
mass. Its mass optimise at $M=1648^{+18}_{-16}$ MeV with 
$\Gamma = 150$ MeV fixed to PDG value. We have tried fits to this peak
with resonances having quantum numbers $\frac{1}{2}^{+}$. We find that 
log likelihood is better for $\frac{1}{2}^{-}$ than $\frac{1}{2}^{+}$ 
by 9.0. With our 4 fitted parameters, the statistical significance
of the peak is $5.8 \sigma$. Our results for $N^*(1650)$ are consistent
with the parameters proposed by PDG.  

A small improvement to the fit is given by including a $J^{P}=\frac{1}{2}^+$
resonance, which optimises at $M=1834^{+46}_{-55}$ MeV and 
$\Gamma = 200$ MeV fixed. The statistical significance of the peak
is only $2.0 \sigma$. we have tried $J^{P}=\frac{1}{2}^{-}$ 
instead $\frac{1}{2}^+$, but the fit is much worse.

\section{Conclusion}

In summary, we have studied the $J/\psi \rightarrow p \bar{p} \eta$ decay
channel, and a PWA analysis is performed on the data. 
There is a definite requirement for a $J^{P}=\frac{1}{2}^-$ component at 
$M = 1540^{+15}_{-17}$ MeV with $\Gamma =178^{+20}_{-22}$ MeV near the 
threshold for $\eta$ production. In addition, there is an obvious 
$J^P=\frac{1}{2}^-$ resonance, with $M = 1648^{+18}_{-16}$ MeV and 
$\Gamma = 150 $ MeV fixed to PDG data. In the higher $p\eta$ 
mass region, there is an evidence of $J^P=\frac{1}{2}^+$ signal around
1800 MeV, we can not get any conclusion for this state due to the low
statistics.

All above analysis is the first step for us to probe $N^*$ baryons at BES.
We will perform detail studies of $N^*$ baryons in the following $J/\psi$
decay channels: $J/\psi \to p \bar{p} \pi^0$, $p \bar{p} \pi^0 \pi^0$, $p
\bar{p}\pi^+ \pi^-$, $p \bar{p}\eta$, $p \bar{p}\omega$ and so on.  

\section{Acknowledgements}
One of the authors, H.B. Li, is grateful to Prof. Leonard Kisslinger
for the helpful discussions. This work
is supported in part by Chinese National Science Foundation under
contract No. 19290401 and 19605007.

\end{document}